# Delta Tensor: Efficient Vector and Tensor Storage in Delta Lake

Zhiwei Bao*, Liu Liao-Liao*, Zhiyu Wu, Yifan Zhou, Dan Fan, Michal Aibin, Yvonne Coady,
Andrew Brownsword†

*Khoury College of Computer Sciences*
*Northeastern University, †Oracle Labs*
Vancouver, Canada
{bao.zhiw, liu.liao, wu.zhiyu, zhou.yifan, fan.da, m.aibin, m.coady}@northeastern.edu, andrew.brownsword@oracle.com

*Abstract*— The exponential growth of artificial intelligence (AI) and machine learning (ML) applications has necessitated the development of efficient storage solutions for vector and tensor data. This paper presents a novel approach for tensor storage in a Lakehouse architecture using Delta Lake. By adopting the multidimensional array storage strategy from array databases and sparse encoding methods to Delta Lake tables, experiments show that this approach has demonstrated notable improvements in both space and time efficiencies when compared to traditional serialization of tensors. These results provide valuable insights for the development and implementation of optimized vector and tensor storage solutions in data-intensive applications, contributing to the evolution of efficient data management practices in AI and ML domains in cloud-native environments.

## I. INTRODUCTION

The increasing importance of vector and tensor storage in the realm of artificial intelligence (AI) and machine learning (ML) is undeniable. Large language models (LLMs), or more broadly, foundation models [1], characterized by their billions of parameters, require massive datasets for training and operation. Their input data come in various forms. These include diverse modalities such as text, voice, images, and videos. In many instances, the volume scales to petabytes. These extensive datasets are used to perform tasks in natural language processing (NLP) and computer vision (CV).

The data often take the form of vectors and tensors. Beyond simple serialization, vector database management systems (VDBMS) have emerged as a crucial solution for storing embedding vectors and their raw feature data. They offer a range of capabilities, including query optimization, transactions, scalability, fault tolerance, and privacy and security for unstructured data [2], [3]. These systems can be broadly classified into two types: "native" systems, specifically developed for vectors, and "extended" systems that incorporate vector indexing and embedding capabilities into existing database systems through extensions.

While numerous studies in VDBMS have contributed to the storage and retrieval of vectors, research dedicated to the efficient storage of both vectors and, their higher-order format, tensors remains limited. Tensors have capabilities that extend beyond those of vectors. They can manage not only embedding vectors but also input datasets and model parameters like weights and biases. It's a unified approach to handling multi-dimensional data. Exploring tensor storage strategies could improve data accessibility and computational efficiency. They are vital for processing complex data types in AI applications. The project aims to bridge this research gap by investigating and developing tensor storage optimization techniques, potentially significantly impacting future developments in the field. Efficient storage techniques for tensors could also be adapted for vectors, providing a general solution.

This project focuses on investigating storage efficiency using an "extended" system, more specifically, the Lakehouse architecture [4]. Unlike vector database systems, this architecture uses Photon [5], a proprietary fine-tuned Spark [6] variant, as its query engine on the Delta Lake storage layer. The choice of the Lakehouse architecture stems from reading complex data like vectors and tensors needs complex non-SQL code. Standard data reading methods like ODBC/JDBC may not offer optimal efficiency. The architecture also benefits from the disaggregation between computation and storage, utilizing cloud object storage, designed with cost and scalability as primary considerations. Tested in Databricks, it has proven its robustness and scalability in handling exabytes of data per day, with the largest instances managing exabyte-scale datasets and billions of objects [7]. The combination of large-scale data processing and cost-effective storage makes the Lakehouse architecture particularly suitable for investigating efficient vector and tensor storage.

---

*Z. Bao and L. Liao-Liao assert joint first authorship.

In the context of "cloud-native" VDBMS, the storage of vectors and tensors also typically employs cloud object storage such as Amazon S3 and Azure Blob Storage. These services offer significant advantages over traditional file systems or data warehouses that use distributed file systems like Hive [8] on HDFS [9]. They free users from maintaining the storage systems and provide benefits including pay-as-you-go billing, economies of scale, and expert management [10].

Despite these advantages, most cloud object storages function primarily as key-value stores. To fully leverage their potential, solutions such as Snowflake [11], an enterprise-ready data warehousing solution for the cloud, have a dedicated storage layer and store their data directly in S3. Similarly, Photon utilizes Delta Lake [7], an open-source ACID table storage layer built on top of cloud object storages. This project has chosen to use Delta Lake as the storage layer with S3 being the underlying object storage service. Photon, being closed-source proprietary software, lacks accessible source code. Therefore, the open-sourced Spark has been chosen as the query engine. While the performance of object storage services can differ among various cloud providers, the findings should still provide valuable insights.

The paper is structured as follows. Section II reviews various techniques and methodologies related to vector and tensor storage formats, along with encoding strategies aimed at enhancing storage efficiency. Section III presents the problem statement. Section IV describes the employed methods and algorithms for tensor storage and optimization. Section V outlines the experimental setup and benchmarks the performance of the proposed methods. Section Section VI summarizes the contributions and findings. Section VII discusses the limitations of current work and identifies future research directions.

## II. Related Work

### A. Multidimensional Array Storage

This project extends investigations beyond vectors to also include tensors, which are typically represented as n-dimensional arrays in most ML frameworks. Baumann *et al.* examined the multi-dimensional data management systems under the "array databases" paradigm. The storage formats they used are tiles [12] or chunks [13] under the assumption that access patterns on arrays are strongly linked to the Euclidean neighborhood of array cells [14]. Thus, by dividing large multi-dimensional arrays into smaller chunks, chunk storage makes data easier to manage. It also allows for quicker and more efficient data retrieval. This is because only the necessary chunk is loaded into memory for processing. Given its advantages, the chunk storage format merits further exploration in the context.

### B. Sparse Encoding Methods

Storing tensors (considering vectors as rank-1 tensors) as n-dimensional arrays can be space-inefficient if the majority of the elements are zeros. A more space-efficient alternative involves encoding the tensors. One such method, referred to as sparse encoding, selectively stores only the non-zero elements along with their positions, resulting in less space usage.

Common sparse encoding methods include Coordinate (COO), Compressed Sparse Row (CSR), Compressed Sparse Column (CSC), and Block Compressed Sparse Row (BCSR). Shahnaz, Usman, and Chughtai *et al.* [15] give their comparative analysis of the storage efficiency of these methods. COO stores each non-zero element along with its row and column indices. Despite being less space-efficient, COO serves as the foundational schema for these methods. CSR and CSC, on the other hand, compress the tensor or vector row-wise or column-wise, respectively, thus saving space by reducing the redundancy in storing positional information. BCSR extends the CSR format to handle blocks of non-zero elements, further optimizing storage for tensors with block-wise non-zero patterns.

These sparse encoding methods have demonstrated effectiveness in compressing sparse matrices as well as vectors without loss of information. However, to apply those methods on tensors, which are multi-dimensional matrices, these methods require further adaptation and generalization.

To adapt sparse encoding methods for tensors, the study conducted by Parker Allen Tew *et al.* [16] presents the Compressed Sparse Fiber (CSF) format as a notable example. This format extends the principles of CSR/CSC, used for sparse matrices, to tensor storage. CSF achieves greater storage efficiency by employing additional layers of index pointer arrays, which compress the tensor's additional dimensions. This method forms a tree-like structure, where each split node, representing a tensor dimension, eliminates the redundancy of index values. The Mode Generic sparse tensor format extends the concept of BCSR to higher-order tensors. It represents a tensor as a sparse collection of dense blocks of any order, with the coordinates of these blocks stored in COO format [17], [18].

### C. Vector Databases

Vector databases serve as specialized platforms designed to store and query vectors and tensors, commonly used for representing unstructured data.

Milvus [19] is an open-source vector database offering rapid and efficient similarity search capabilities for high-di-mensional unstructured data. This is achieved through the utilization of advanced indexing and search algorithms. Milvus adopts a storage approach where vectors are continuously stored and arranged in a columnar format, resulting in reduced storage overhead and an enhanced cache hit rate. Additionally, Milvus adopts quantization techniques such as product quantization (PQ) or scalar quantization (SQ) to compress vectors, leading to further reductions in storage space and increased query speed.

Deep Lake [20] is a novel lakehouse designed for deep learning applications. Deep Lake can store complex data types, such as images, videos, annotations, and tabular data, in the form of tensors. Tensors are stored in chunks using a columnar storage format as arrays with types and shapes. Each tensor comes equipped with an index map for sample retrieval, and they can be grouped and nested to illustrate relationships. Deep Lake also possesses the ability to track changes in the dataset's schema and content over time.

### D. Limitations and/or Gaps

Existing research on VDBMS primarily addresses aspects such as query optimization, transactions, scalability, fault tolerance, and privacy and security. However, the specialized subject of vector storage has not been thoroughly explored. This project investigates vector storage by examining it in its more generalized form - tensors. The high dimensionality and properties of tensors introduce specific challenges as the shape and dimensions can vary arbitrarily.

Despite extensive research [15], [16] in sparse matrix encoding, there remains a notable gap in generalizing these techniques for high-dimensional applications. Many studies focus on lower-dimensional vectors and matrices. They do not address the challenges of data structures with higher dimensions. Furthermore, these studies use traditional file systems. Their findings may not apply to cloud-native storage. The research intends to fill this gap by adapting existing storage formats and encoding methods to meet the specific needs of a cloud-native storage environment, specifically, the delta lake table storage layer on top of S3. This project aims to focus on generalizing these methods for high-dimensional data, exploring their practicality and effectiveness in high-dimensional, large-scale data environments.

## III. PROBLEM STATEMENT

### A. Terminology

Currently, tensor data are usually stored as binary serialization blob files in databases. In most of the cases, tensor tends to be sparse. This naive serialization method does not optimally utilize storage space. Given the substantial data volumes that can scale up to petabytes, this method results in significant and unnecessary space waste. This project aims to apply existing storage and encoding techniques to tensors to improve efficiency and performance, which forms the central focus of the research.

Tensor is the generalized form of vector and matrix. In ML tasks, tensors are usually represented as multidimensional arrays. A vector, also known as a one-dimensional tensor, is commonly denoted as $\boldsymbol{a} = [a_1, a_2, ..., a_n]$, where $n$ represents the size of the vector. A two-dimensional tensor, a matrix, is represented as

$$A = \begin{bmatrix} a_{11} & a_{12} & ... & a_{1n} \\ a_{21} & a_{22} & ... & a_{2n} \\ \vdots & \vdots & \ddots & \vdots \\ a_{m1} & a_{m2} & ... & a_{mn} \end{bmatrix} \quad (1)$$

where $m$ and $n$ denote the size of the matrix, and each row and column can also be viewed as a vector.

A tensor can have arbitrary dimensions. An N-dimensional tensor $X$ with dimension $D = (d_1, d_2, ..., d_N)$ can be represented as

$X = \{x_{i^1, i^2, ..., i^N} \mid i^j \in [1, d_j] \text{ for } j = 1, 2, ..., N\}$ where $x_{i^1, i^2, ..., i^N}$ represents an element in the tensor $X$, and $i^j$ are the **indices** of that element in dimension $d_j$.

The slice operation on a tensor $X$ selects a subset of $X$ by fixing one or more indices to a specific value. If we denote the slice indices as $S = (s^1, s^2, ..., s^M)$, then a slice of $X$

$$X_S = \{x_{i^1, i^2, ..., i^N} \mid i^j = s^j \text{ for } j \in [1, M] \\ \cup \, i^j \in [1, d_j] \text{ for } j \in [M+1, N]\} \quad (2)$$

where $M$ is less than or equal to $N$. For example, the equivalent of `X[0:100,:,:,:]` in Python `numpy` notation could be represented as

$$X_s = \{x_{i^1, i^2, i^3, i^4} \mid i^1 \in [1, 100] \\ \cup \, i^j \in [1, d_j] \text{ for } j \in [2, 4]\}. \quad (3)$$

For simplicity, the part that collects all the indices in a dimension $d_j$ ($\cup \, i^j \in [1, d_j]$ for $j \in [M+1, N]$) is ommited, thus the equivalents of (2) and (3) are:

$$\begin{aligned} X_S &= \{x_{i^1, i^2, ..., i^N} \mid i^j = s^j \text{ for } j \in [1, M]\} \\ X_s &= \{x_{i^1, i^2, i^3, i^4} \mid i^1 \in [1, 100]\} \end{aligned} \quad (4)$$

Fibers serve as the higher-order analogs to rows and columns in matrices. In the context of matrices, a column corresponds to a mode-1 fiber, while a row represents a mode-2 fiber. In the case of third-order tensors, there are column, row, and tube fibers, denoted as $X_{:jk}$, $X_{i:k}$, and $X_{ij:}$ respectively [21]. If a slice operation covers all the indices in some dimensions, the result can be referred to as the corresponding fiber such as $X_{i::}$, $X_{:j:}$, and $X_{::k}$. For convenience, this notation is used to denote both the slice operation and the result of such a slice operation of the tensor. For the previous example, `X[0:100,:,:,:]` can be represented as $X_{[1:100]:::}$.

An encoding method, denoted as $F$, converts $X$ into a different format $X_{\text{encode}}$. This transformation is mathematically represented as:

$$X_{\text{encode}} = F(X) \tag{5}$$

Additionally, this encoding method is accompanied by a corresponding inverse operation, functioning as the decoding process, which can be expressed as:

$$X = F^{-1}(X_{\text{encode}}) \tag{6}$$

### B. Performance Metrics

The evaluation of the efficiency of proposed methods can be categorized into two main aspects: storage space efficiency and processing time efficiency.

To elaborate, concerning storage space, the compression ratio for a tensor is defined as follows:

$$C_r = \frac{S_{\text{encode}}}{S_{\text{binary}}} \tag{7}$$

where $S_{\text{binary}}$ represents the original space occupied by a tensor when serialized as binary file, while $S_{\text{encode}}$ denotes the space required after the tensor is encoded.

Regarding time efficiency, it can be broken down into read efficiency and write efficiency. Both reading and writing involve considerations of the time taken for input/output operations, indicated as $t_{\text{ser}}$ for serialization during writing and $t_{\text{des}}$ for deserialization during reading. Additionally, it is necessary to account for the time required for encoding and decoding individual tensors, represented by $t_{\text{en}}(X)$ and $t_{\text{de}}(X)$ respectively, and defined as follows:

$$\begin{aligned} t_{\text{en}}(X) &= Elapsed(F(X)) \\ t_{\text{de}}(X_{\text{encode}}) &= Elapsed(F^{-1}(X_{\text{encode}})) \end{aligned} \tag{8}$$

For write efficiency, it comprises the time spent on data serialization, along with the time taken for encoding, defined as follows:

$$t_{\text{write}} = t_{\text{ser}} + t_{\text{en}}(X) \tag{9}$$

The efficiency of reading involves the time dedicated to data deserialization and decoding for each tensor. Read operations can be further categorized into two types: read the entire tensor or a slice of the tensor. The time spent for the above two read operations is denoted as $t_{\text{read\_tensor}}$, and $t_{\text{read\_slice}}$ respectively, and defined as follows:

$$\begin{aligned} t_{\text{read\_tensor}} &= t_{\text{des}} + t_{\text{de}}(X_{\text{encode}}) \\ t_{\text{read\_slice}} &= t_{\text{des}} + t_{\text{de}}(X_{\text{S\_encode}}) \end{aligned} \tag{10}$$

where $X_{\text{S\_encode}}$ denotes the slice operation of an encoded tensor. It's worth mentioning that the slice operation may occur either during the deserialization or decoding operation, depending on the specific implementation of the storage method.

This paper presents the design and implementation of five distinct tensor storage methods, with a focus on evaluating their compression ratios and read/write performance using direct (de)serialization as the baseline of the comparisons. The novel contribution of this paper lies in enabling efficient tensor storage within cloud object storage, an area that has not been thoroughly investigated before.

## IV. METHODOLOGY

In this section, a distinction is made between two categories of tensors: general tensors and sparse tensors. This differentiation aids in determining whether to apply *sparse* encoding methods. In the specific subsections detailing specific encoding methods, a separation is made between the encoding algorithm and the underlying storage formats. This is necessary as tensor data structures change based on the encoding method used.

The storage methods discussed in the following subsections will eventually persist tensors in the Parquet file format. Meanwhile, Delta Lake serves as a storage framework built on top of Parquet, offering features such as ACID transactions, time travel, and various optimizations [7]. In online transaction processing, which involves frequent data writing, the data are usually structured in a row-based format. Conversely, for online analytical processing, where data reading predominates, a columnar format is often preferred. Parquet, however, utilizes a hybrid format that leverages the advantages of both row-based and columnar formats [22].

### A. Flattened Tensor Storage Format

Tensors can be treated as multidimensional arrays. In the multidimensional array storage format, array elements are typically organized into fixed-size *chunks*. These chunks are hyper-rectangles, with each dimension's index range defined [23]. When this storage format is applied to tensors, two crucial parameters need consideration: the chunk *size* and the chunk *shape*. For optimal performance in the file system, the chunk size is generally matched with the file system block size, which is the smallest read and write unit. The optimal chunk shape, on the other hand, depends on the access pattern. Ideally, the chunk shape should be chosen to minimize the average number of file system block fetches for a given access pattern.

However, these practices for determining optimal chunk size and shape cannot be applied in all contexts. For instance, when tensors are stored in a cloud object storage, the block size of the object storage isn't exposed to users. Regarding the chunk shape, Sarawagi et al. suggested that if the access pattern isn't given, the DBMS can use a default chunk shape and monitor the access statistics for re-chunking later [13]. This approach, however, is infeasible in projects operating only on the storage layer, where user statistics are not available.

*1) Algorithm:*

The Flattened Tensor Storage Format (FTSF) is proposed as a solution for *general* tensors. Instead of storing tensors in files using the multidimensional array storage format, they are stored in a Delta Lake table. The tensors are still chunked, but the chunk is a lower-ranked tensor from the original tensors. Given a chunk dimension $D^c$, the result of applying the encoding method $F(X, D^c)$ can be represented as:

$$X_{\text{encode}} = F(X, D^c)$$
$$= \left\{ X_{s_i} \mid s_i \text{ for } i = [1, d_1 \cdot d_2 \cdot ... \cdot d_{N-D^c}] \right\}$$

where $N - D^c$ is the dimension of the resulting chunked tensor in which the last $D^c$ dimensions are "merged" (they cannot be directly accessed after the chunking). And $s_i$ contains the indices of first $d_1...d_{N-D^c}$ dimensions.

This equation represents the function $F$ that takes a tensor $X$ and a chunk dimension $D^c$, and returns an *array* of chunks where each chunk is a fiber $X_{s_i}$ with dimension flattened to $D^c$.

*2) Storage format:*

*Metadata*

| dim_count | dimensions | chunk_dim_count | ... |
|---|---|---|---|
| 4 | [24, 3, 1024, 1024] | 3 | ... |

Figure 1: The metadata columns

| id | chunk (BINARY) | Metadata* |
|---|---|---|
| 6e368... | 1 | |
| 6e368... | 2 | |
| | ... | ... |
| 6e368... | 24 | |
| 1a234... | 1 | |
| | ... | |

a) The FTSF table

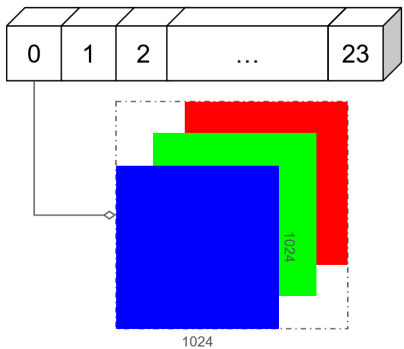

b) A(24, 3, 1024, 1024) tensor

Figure 2: FTSF with 3D chunks

Chunks afterward are stored in the Delta Lake table, and FTSF will create a row for each chunk with the metadata. Figure 2 (b) provides a conceptual view of a 4D tensor of a 24-frame sequence of 1024x1024 RGB images chunked by their 3D tensors. Figure 2 (a) shows an example table with metadata in Figure 1.

In FTSF, the id: STRING serves as a unique identifier for each tensor. The chunk: BINARY is a binary chunk with the semantics of a tensor of a user-specified rank. In this example, it's a 3D tensor. dim_count: INT represents the number of dimensions of the original tensor, while dimensions: ARRAY <INT> is an integer array of size dim_count, storing the size of each dimension of the tensor. chunk_dim_count is the dimensionn $D^c$ of the chunk.

When the chunk dimension changes, the metadata adjusts accordingly. Figure 3 illustrates the metadata columns when the same tensors are flattened as 2D chunks. Notably, the Delta Lake table uses the Apache Parquet format [22] for data organization, which employs dictionary encoding. Thus, even though the same metadata recurs across multiple rows, it compresses efficiently. Delta Lake also allows schema evolution, enabling users to add their own metadata columns for customization. In this subsection, tensors are stored in plain serialization. The *sparse* encoding methods introduced later will also leverage the schema evolution feature to apply sparse encoding methods on tensors by modifying the metadata columns.

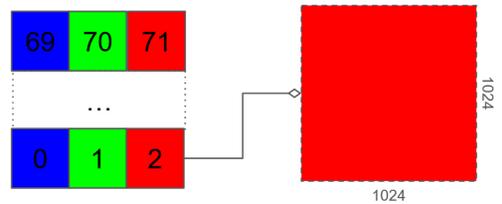

a) The tensor view when flatten as 2D tensors

*Metadata*

| dim_count | dimensions | chunk_dim_count | ... |
|---|---|---|---|
| 4 | [24, 3, 1024, 1024] | 2 | ... |

b) The metadata columns when flatten as 2D tensors

Figure 3: FTSF with 2D chunks

### B. Sparse Tensor Characteristics and Encoding

Sparse tensors primarily contain zero elements. The Formidable Repository of Open Sparse Tensors and Tools (FROSTT), a collection of publicly accessible sparse tensor datasets and tools [24], doesn't set a criterion for classifying a tensor as sparse. However, almost all tensor datasets in FROSTT have non-zero elements constituting less than 10% of the total tensor elements. This trait of sparse tensors al-lows for storage optimization with *sparse* encoding methods, as it eliminates the need to store numerous zero elements. For general tensors, where most elements are non-zero, these sparse encoding methods are not applied, as the information entropy, i.e., the amount of "information," "randomness," or "uncertainty" is inherently high.

The determination of the sparsity threshold depends on the specific application and the balance between time complexity and storage efficiency. Setting a higher threshold may classify more tensors as sparse, which could potentially improve storage efficiency. However, this could simultaneously increase the computational load involved in encoding and decoding. On the other hand, a lower threshold would categorize fewer tensors as sparse, thus sparing the computation but possibly giving up storage efficiency. For the purposes of this research, a rule of thumb: the 10% threshold is used to decide whether a tensor is categorized as sparse.

Sparse tensors can be managed efficiently using four methods proposed below: Coordinate Encoding (COO), Compressed Sparse Row/Compressed Sparse Column (CSR/CSC), Compressed Sparse Fiber (CSF), and Block Sparse Generic Storage (BSGS). In addition to COO, which serves as a foundational and flexible storage method for sparse tensors, the other techniques, built upon COO, can be categorized into two groups: Encoding before Partitioning, and Partitioning before Encoding.

The first group includes CSR/CSC and CSF formats, which focus on applying encoding methods to the tensor first, resulting in encoded arrays. Since these arrays can be large, they are partitioned into separate chunks, thereby significantly enhancing tensor writing speed by dramatically reducing data size.

In contrast, the second group, including BSGS, initially breaks down the tensor into smaller blocks before applying encoding methods to each block individually. This approach enables tensor slicing functionality before decoding, eliminating the necessity to read or decode the entire tensor when only a specific slice is needed. Consequently, this method accelerates accessing tensor slices.

### C. Coordinate Encoding

*1) Algorithm:*

COO offers an efficient approach for storing sparse tensors. This method involves storing all non-zero elements (abbreviated as nnz) of a tensor along with their respective coordinates in separate vectors. The space requirement for this method is $O(nnz)$ for the values and $O(m \cdot nnz)$ for the index, where $m$ represents the number of dimensions. The key advantage of COO encoding is its ability to save substantial storage space by excluding all zero values and their indices, thereby making it an effective solution for handling sparse tensors. An simple example of COO is provided in Figure 4.

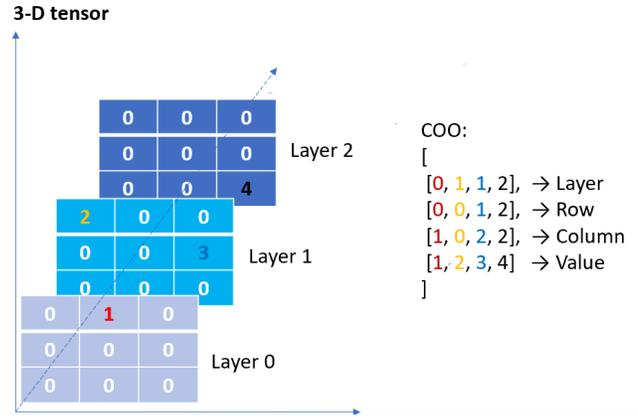

Figure 4: The Coordinate Encoding Method

Decoding a sparse COO-formatted tensor involves a process of reconstructing the original tensor from the stored non-zero elements and their corresponding coordinates. However, this process can occasionally introduce inaccuracies due to the absence of explicit information about the actual size or shape of the original tensor. To mitigate this issue, an additional list is utilized to store the shape of the tensor. This supplementary information assists in the accurate reconstruction of the tensor, ensuring that the dimensions of the decoded tensor align with those of the original tensor.

*2) Storage Format:*

| id | layout | dense_shape | indices | value |
|---|---|---|---|---|
| 12cac.. | COO | [3, 3, 3] | [0, 0, 1] | 1.0 |
| 12cac.. | COO | [3, 3, 3] | [1, 0, 0] | 2.0 |
| 12cac.. | COO | [3, 3, 3] | [1, 1, 2] | 3.0 |
| 12cac.. | COO | [3, 3, 3] | [2, 2, 2] | 4.0 |

Figure 5: The Coordinate Storage Format

In accordance with the strategy to enhance the reliability and accuracy of COO, the storage schema includes three additional fields: `id`, `layout`, and `dense_shape`.

Figure 5 demonstrates the storage format of COO in Delta Lake table. The `id` column assigns a common identifier to all elements of a specific tensor, providing a unified reference to the entire tensor. The `layout` column indicates the encoding method used, and the `dense_shape` column captures the shape of original tensor. The `indices` and `value` columns record the position and value of each non-zero element, respectively. These measures collectively bolster the robustness of COO storage format in encoding and decoding sparse tensors.

### D. Compressed Sparse Row & Compressed Sparse Column

*1) Algorithm:*

This section introduces an approach to handle sparse tensors with CSR/CSC encoding formats, which are traditionally used for sparse matrix representations. This method reshapes sparse tensors into 2D matrices while preserving their inherent sparsity. Once the 2D matrices are obtained, CSR/CSC can be applied on them, optimizing storage and computation for tensor-based data. The CSR algorithm uses three arrays to efficiently represent the non-zero elements, their column indices, and row pointers indicating the start of each row. Conversely, the CSC algorithm applies the same concept to columns, using three arrays for non-zero elements, their row indices, and column pointers to denote the beginning of each column.

The effectiveness of this approach resides in its ability to maintain the sparsity of the original tensors through the reshaping process while reaping benefits from efficient storage. Since only non-zero elements and their positional information are stored, the space required to store the sparse tensors can be significantly reduced. Furthermore, this method allows the utilization of optimized libraries and algorithms designed for sparse matrix operations, thereby improving the execution speed for tensor-based computations.

Additionally, this approach offers flexibility in processing sparse tensors, supporting both row-wise and column-wise operations. This versatility proves crucial in multidimensional data analysis, as the method of data manipulation can significantly affect the outcome of the computations.

2) *Storage Format:*

For CSR/CSC, since only 2-d matrices are supported, tensors are converted to 2-d matrices first. For matrices beyond two dimensions, a flattening procedure is applied to the indices before conversion. Before conversion, the original shape of the tensor is recorded as `dense_shape`; after conversion, the flattened shape of the 2-d matrix is recorded as `falttened_shape`. These 2 shapes help restore the tensor when needed.

After flattening, CSR/CSC is applied on the 2-d matrix, resulting in three arrays. For CSR: `value`, represents the non-zero values in the matrix; `col_indices`, represents the column index of the non-zero values; `crow_indices`, represents the index in `value` and `col_indices` where the given row starts. For CSC: `value`, represents the non-zero values in the matrix; `row_indices`, represents the row index of the non-zero values; `ccol_indices`, represents the index in `value` and `row_indices` where the given column starts. To restore the original tensor, flatten process is reversed given the original shape of the tensor.

The internal storage layout for a CSR/CSC-formatted tensor is recorded as follows:
- `id`: the tensor ID
- `layout`: the storage type, which is `CSR` or `CSC` in this case
- `flattened_shape`: the shape after tensor reshape/flattening, which is used to restore the 2-d matrix back to the original tensor
- `dense_shape`: the original shape of the tensor, used in the restore process
- `crow_indices/ccol_indices`,`col_indices/row_indices`, and `value`: the 3 essential arrays representing the non-zero elements and their locations in the 2-d matrix

*E. Compressed Sparse Fiber*

This section introduces the CSF storage format, an extension and optimization of the traditional CSC/CSR formats, specifically designed for higher-order tensors. CSF is a general approach to handle sparse tensors with dimensions greater than two, making it a versatile choice for multidimensional sparse data storage.

1) *Algorithm:*

For higher-order tensors, CSF format organizes the elements in tensors into a hierarchical framework of fibers, each marked by a distinct dimensionality. This methodical compression initiates from the tensor's non-zero elements, and then arranges them into a tree-like architecture where each layer represents a fiber with unique indices and pointers.

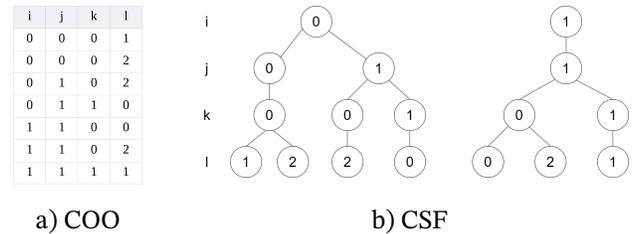

a) COO  b) CSF

Figure 6: COO and CSF formats for a four-dimension tensor.

Figure 6 illustrates an example of applying CSF format on a tensor originally represented in COO format. Each level of the tree corresponds to a dimension of the tensor, with nodes representing non-zero elements and their index within that dimension. Edges connect elements along dimensions, with leaves holding the actual values. When a node is split into its lower-level subtrees or leaf nodes, duplicate index values are removed.

2) *Storage Format:*

Similar to CSR/CSC, in actual implementation, the conceptual tree structure is packed into arrays. For each dimension, two arrays are added: fiber pointers (`fptrs`) and fiber indices (`fids`). Fptrs indicates how to slice the next dimension, while `fids` stores what unique indices are present in that dimension. In addition, all non-zero values of the tensor are packed into a separate array.

To build the CSF format for a tensor, initially, a root fiber is defined, encompassing the first dimension of the tensor.

This fiber acts as the entry point to recursively decompose the tensor into its constituent fibers. A breadth-first search (BFS) algorithm iterates through the tensor's dimensions, systematically compressing each fiber. For each dimension, a counter is maintained to track the occurrence of indices, which helps in creating the fiber pointers and indices. These pointers and indices are essential for navigating the compressed tensor structure. When the tensor is being traversed, the fiber pointers and indices are dynamically adjusted based on the encountered non-zero elements, ensuring an efficient compression that reflects the tensor's sparsity pattern.

The internal storage layout for a CSF-formatted tensor is structured as follows: Each tensor is identified by a unique ID that concatenates a prefix, the tensor's dimensionality, and a randomly generated ID string. The layout information, dense shape of the tensor, non-chunked indices and non-chunked pointers for the first two tensor dimensions are stored once per tensor. For values, as well as indices and pointers in the remaining dimensions, storing each of them in a single array would be time and space inefficient. Instead, the array representations of the above information are chunked, with each chunk given its own unique sub-identifier and stored alongside its respective metadata in Delta lake.

The non-chunked data includes:
- `id`: Unique tensor ID for CSF format.
- `layout`: CSF
- `dense_shape`: The full dimensional shape of the tensor.
- `fptr_zero` and `fptr_one`: Fiber pointers for the tensor's first two dimensions.
- `fid_zero` and `fid_one`: Fiber unique indices for the first two dimensions.

The chunked data is divided into manageable sizes, each containing:
- Subsets of indices and pointers for dimensions beyond the first two.
- Corresponding metadata to reconstruct the location and structure of data within each chunk.

To read and reconstruct the entire tensor, the program first identifies the tensor's unique ID and retrieves the associated chunked and non-chunked data. Then, it aggregates the indices recursively for each dimension to reconstruct the tensor's structure.

The program could also reconstruct a sub-tensor slice, by performing slicing operations directly on the indexed data with values in a specific range.

In summary, the CSF format provides a robust solution for writing and reading sparse tensors. It reduces storage costs by compressing duplicate indices and enhances access speeds through optimized data storage mechanisms like chunk storage.

### F. Block Sparse Generic Storage

This section introduces BSGS, which builds upon the concepts of BCSR and Mode Generic tensor format. It's a derived format that incorporates customizations tailored to fit specific physical storage arrangements.

1) *Algorithm:*

The proposed algorithm can be described for both reading and writing as follows:

For writing a tensor: The tensor is first converted into Mode Generic sparse tensor format with an appropriate block shape. This sparse tensor format is then stored with a customized storage layout in Parquet file format using Delta Lake.

For reading a tensor: The sparse tensor format is retrieved using the provided tensor ID and slicing information. Subsequently, the tensor with its original shape or sliced shape is reconstructed and output.

The BCSR extends the concept of CSR by partitioning the tensor into smaller groups and storing the dense blocks of non-zero elements along with their respective block indices.

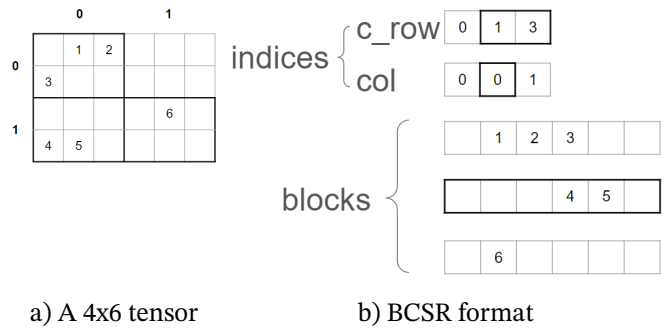

a) A 4x6 tensor    b) BCSR format
Figure 7: BCSR format for a 2d tensor.

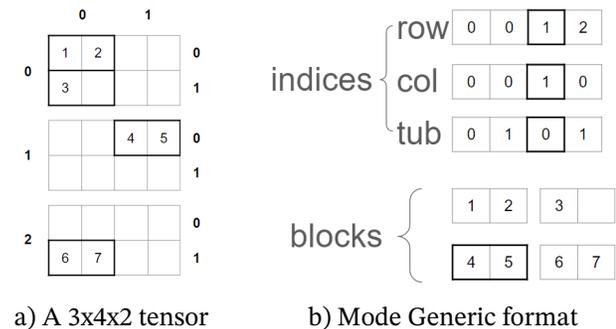

a) A 3x4x2 tensor    b) Mode Generic format
Figure 8: Mode Generic format for a 3d tensor.

In a 4 x 6 tensor, as illustrated in Figure 7, BCSR would take the block size of 2 x 3 as input, dividing the tensor into four blocks. Each non-zero block is then gathered along with its corresponding row and column indices (represented as "col" in the figure). Duplicate row values are compressed, denoted as "c_row" in the figure. For instance, in Figure 7, the coordinates of the last two rows are duplicated for the final two blocks.

The Mode Generic sprase tensor format generalized the idea of BCSR into higher order tensors. It divides the tensor into blocks of any order and store the non-zero dense blocks along with their respective block indices.

Given the 3 x 4 x 2 tensor in Figure 8, a block size of 1 x 2 is selected to partition the tensor. Non-zero blocks are collected alongside their corresponding row, column, and tube indices, as denoted by "row", "col", and "tub" respectively in the figure.

The choice of block size plays a crucial role in the performance of both writing and reading tensors, as well as the space cost of tensor storage. If the block size is too large and matches the dense shape of the tensor, it results in wastage of space as many zero values are stored. Conversely, if the block size is too small, or in the extreme case where each element is treated as a separate block, similar to the COO format, it can lead to increased I/O operations, losing the benefit of sequential data scanning provided by larger blocks.

2) *Storage Format:*

| id | dense_shape | block_shape | indices | values |
|---|---|---|---|---|
| 1, 4 | [3, 4, 2], 4 | [2, 1], 4 | [0, 0, 0] | [1, 2] |
| | | | [0, 0, 1] | [3, 0] |
| | | | [1, 1, 0] | [4, 5] |
| | | | [2, 0, 1] | [6, 7] |

Figure 9: Internal storage layout for the tensor in Figure 8.

The internal storage layout of the Mode Generic tensor format is structured as follows. Each block in the sparse tensor is flattened into a vector, which is then stored in an internal table along with its block indices, block shape, and the dense shape of the original tensor. Additionally, a unique ID is generated for the entire tensor. The layout resembles that shown in Figure 9. Notably, given the benefit of columar storage, the column compression is applied to the duplicate values like id, dense_shape, and block_shape. Here, the value 4 after the comma signifies that there are 4 consecutive identical values.

To read an entire tensor, the program first locates the given ID, scans, and populates all the shape records, indices, and values to reconstruct the original tensor. Tensor slicing operations can also be easily applied in this storage format. For instance, if a user needs to obtain the first row of a tensor with ID 1, denoted as $X_{[1]::}$, the program follows these steps:

1. Scan the internal tensor table to locate ID 1.
2. Retrieve the dense shape of the original tensor to form the shape of the slice, which in this case is 1 x 4 x 2.
3. Obtain the block shape to reshape the values into blocks in the subsequent step.
4. Scan the indices column to filter the rows that meet the slicing requirement and retrieve the corresponding values.
5. Reshape the values into blocks using the block shape and reconstruct the slice using indices and slice shape.

With this layout, users can apply slicing operations to access a large tensor without needing to scan and construct the entire tensor. In scenarios where the block exhibits sparsity, additional encoding methods such as COO can be applied to each block to conserve storage space further.

In summary, the proposed method offers several benefits and contributions. Firstly, it utilizes the Mode Generic tensor format to reduce the space cost of sparse tensors, while leveraging column compression in columnar storage to minimize the storage space by eliminating redundant data. Additionally, this approach facilitates fast access speeds through the sequential scanning advantages of block storage. Moreover, it provides efficient slicing operations, enabling quick access to specific parts of the tensor without the necessity of scanning the entire tensor.

## V. EXPERIMENTS

This section presents experiments evaluating the performance metrics defined in the Section III. The datasets are first loaded into a spark cluster of 2 Intel® Xeon® Gold 5215 processors with 128GB RAM. The network bandwidth is 1 Gbps. As suggested in Section IV Methodology, two scenarios are considered:

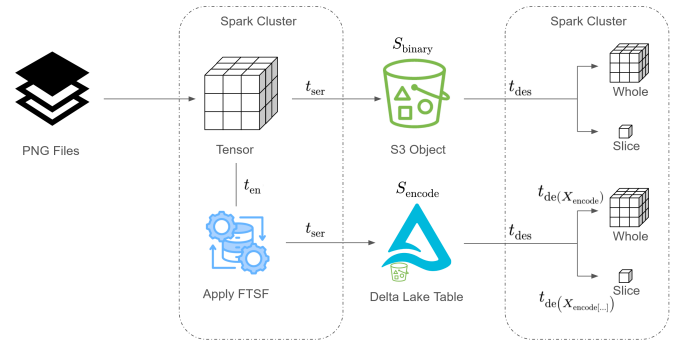

Figure 10: The evaluation plan for storing a dense tensor

1. Storing a *dense* tensor (general tensor that has much more than 10% non-zero elements) derived from the Flickr-Faces-HQ Dataset (FFHQ) [25]. The FFHQ dataset contains 70,000 high-quality 1024x1024 resolution PNG images of human faces. Due to the hardware limitation, a subset of 5,000 images are chosen for the experiment. These images are processed into a tensor of dimension (5000, 3, 1024, 1024). The proposed FTSF is applied in this scenario with no sparse encoding method.

2. Storing a *sparse* tensor constructed from the Uber Pickups Dataset [26]. This dataset includes Uber pickup data in New York City from April to August 2014. The resulting tensor has dimensions of `(183, 24, 1140, 1717)` with 3,309,490 non-zero elements. When represented in dense format, this tensor occupies 8,596,812,960 elements, highlighting its sparsity with only 0.038% of non-zero elements. Given that the original data is already in COO format, there's no requirement to encode this dataset to COO again. However, there are encoding and decoding methods available for converting between COO and various other formats including CSR, CSF, and BSGS. The methods for sparse tensor storage including COO, CSR, CSF, and BSGS from the methodology section are applied in this scenario.

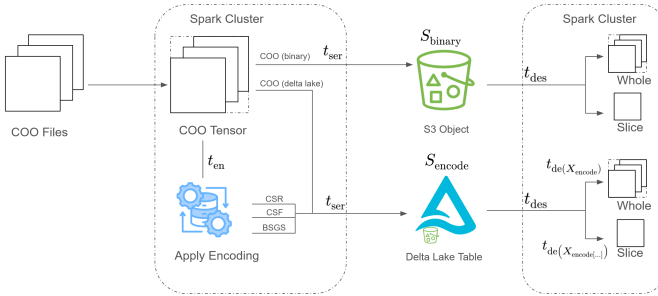

Figure 11: The evaluation plan for storing a sparse tensor

The dense tensor is in the format of `numpy.array` [27] when loaded into memory and serialized as an S3 object. In contrast, the sparse tensor is represented in memory as `torch.sparse_coo_tensor` and stored in S3 using the PyTorch PT file format [28]. Performance metrics defined in Section III.B are evaluated.

A. *Experimental results on dense tensors*

|  | Storage Size (GB) | Write Tensor (s) | Read Tensor (s) | Read Slice (s) |
| --- | --- | --- | --- | --- |
| Binary | 14.6 | 135.69 | 379.51 | 494.33 |
| FTSF | 13.3 | 251.77 | 474.51 | 49.24 |
| $\Delta$ | $-8.90\%$ | $85.52\%$ | $25.02\%$ | $-90.04\%$ |

Figure 12: Performance of different operations on the FFHQ dataset. For the read slice operation, a random fiber of $X_{[1:100]:::}$ was fetched (100 images).

The design decision to apply FTSF to general tensors stems from the practical applications of tensors. Tensors mostly serve as the training data. Although Stochastic Gradient Descent (SGD) factors into this consideration, the reality of training typically taking place on GPUs with limited VRAM makes it often unfeasible to load an entire dataset into memory. The use of batches becomes necessary. As such, fetching a slice of the tensor is a more common use case than retrieving the whole tensor. By using a chunking strategy for the tensor, we can efficiently fetch only the specific chunks that have a particular batch of the dataset when needed.

The results shown in Figure 12 demonstrate that FTSF yields promising outcomes when reading a slice of the tensor, a common use case. This is anticipated as only relevant chunk rows are fetched, while in contrast, for binary, the entire tensor must be retrieved to read the slice.

An overhead of $85.52\%$ for writing the tensor seems concerning. However, this should not be seen as a significant concern for two reasons. First, in practical scenarios, read operations disproportionately surpass write operations. Compared to the $90.04\%$ reduction in slice reading, this overhead is a reasonable trade-off. Secondly, 82.413 seconds (accounting for $60.73\%$) were spent utilizing a Python for loop to create the Spark Resilient Distributed Dataset (RDD) [6]. This suggests substantial room for optimization. For reading the tensor, the overhead $25.02\%$ comes from Spark scheduling, as there's a similar overhead of $85.52\% - 60.73\% = 24.79\%$ (subtracting the time spent on creating RDDs) in the read operation.

An interesting aspect is the compression rate, $C_r = \frac{S_{\text{encode}}}{S_{\text{binary}}} = 91.09\%$, even though no actual compression encoding method was applied. Chunks are serialized as bytes using the `numpy.save` method. Despite the additional metadata in other columns, a size reduction of $8.90\%$ is still realized.

B. *Experimental results on sparse tensors*

To assess performance, methods for sparse tensor storage including COO, CSR, CSF, and BSGS from the methodology section are employed for benchmarking against the current PyTorch PT file format. Due to the interchangeable nature of CSR and CSC, only CSR is evaluated as a representative for performance benchmarking.

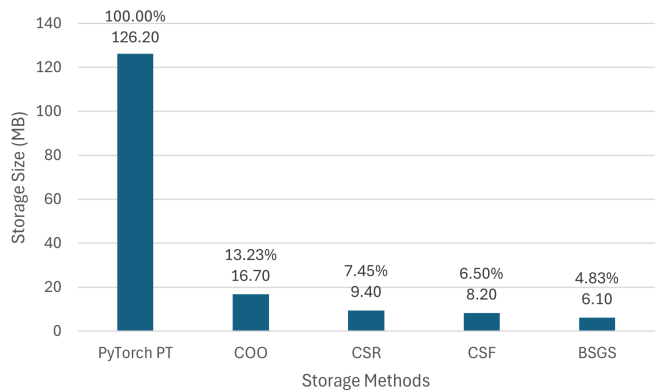

Figure 13: Storage sizes of Uber Pickiups dataset using different storage methods

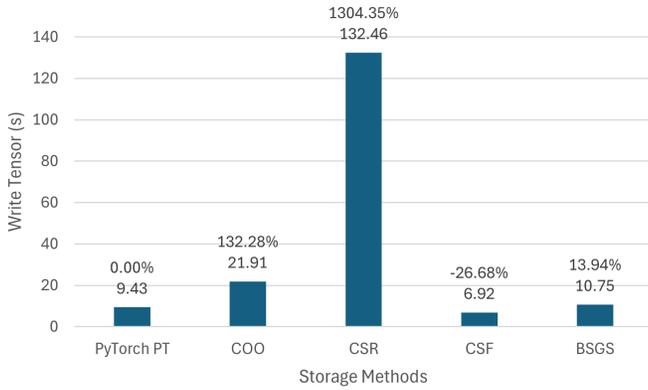

Figure 14: Performance of writing Uber Pickiups dataset using different storage methods

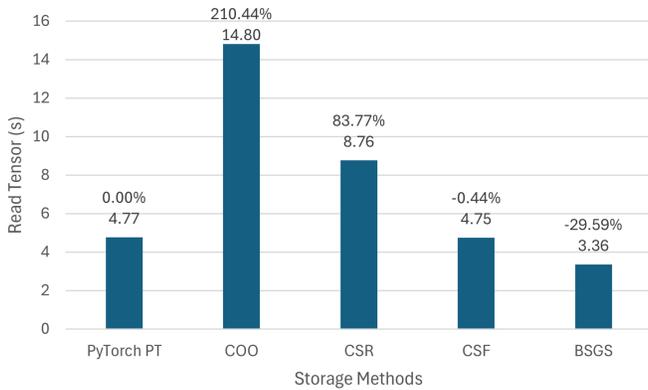

Figure 15: Performance of reading entire Uber Pickiups dataset using different storage methods

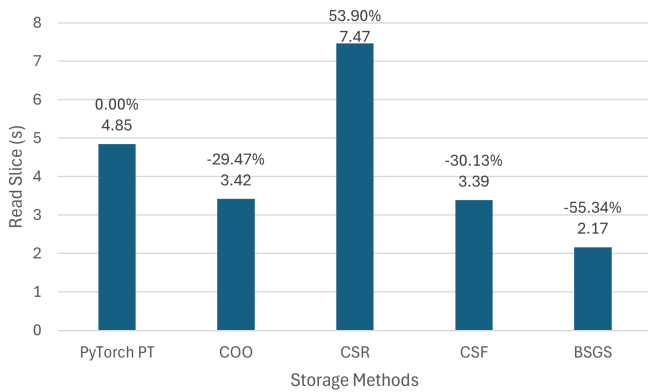

Figure 16: Performance of reading slices of Uber Pickiups dataset using different storage methods

As depicted in Figure 13, in terms of the storage size, all proposed sparse tensor storage methods outperform the PT file format, with compression rates $C_r$ less than 13.23% of the original dataset. Among these, BSGS achieves the best $C_r$ at 4.83%.

Time taken to write and read sparse tensors, denoted in Section III as $t_{\text{write}}$, $t_{\text{read\_tensor}}$, $t_{\text{read\_slice}}$ are crucial metrics for evaluation. To mitigate the impact of network conditions, each of these operations is repeated 100 times, and the average processing time is calculated as the final result. For the slice reading operation, a slice of the first dimension, denoted as $X_{[i]:::}$, where $i$ represents the index of the first dimension of the Uber Pickups dataset ranging from 0 to 183, is considered.

As shown in Figure 14, CSF and BSGS demonstrate comparable performance in terms of writing tensor, with CSF being the most efficient among all others, taking 26.68% less time compared to PT.

Figure 15 and Figure 16 illustrate the efficiency of reading the entire tensor and its slices, respectively. In terms of reading the entire tensor, both CSF and BSGS show comparable read efficiency, with BSGS being the most efficient, taking 29.59% less time than PT. When it comes to reading slices of the tensor, COO, CSF, and BSGS all outperform the PT file format, with BSGS being the most efficient, taking 55.34% less time compared to PT.

In summary, all the proposed methods outperform the PT file format in terms of data storage size, with considerable compression rates. Among these, CSF and BSGS stand out as recommended storage solutions for sparse tensors. CSF excels in writing performance, while BSGS demonstrates superior compression rates and reading efficiency, particularly for partial reads such as reading tensor slices.

## VI. CONCLUSION

In conclusion, this research has explored and evaluated diverse methods for optimizing tensor storage in cloud object storage, an area that has seen limited exploration to date. The investigation has covered the storage of both general and sparse tensors, with the latter being the center as they are prevalent in machine learning applications. The proposed techniques have been adapted from existing storage formats and encoding methods, tailored to the Delta Lake table.

The findings indicate that all the proposed methods improved the space efficiency compared with the serialization of the `numpy.array` or PyTorch PT. Notably, the CSF and BSGS techniques have emerged as the most efficient options. CSF stands out for its exceptional writing performance, while BSGS showcases superior compression rates and reading efficiency, particularly for partial reads.

This study contributes to the ongoing advancements in data management systems in a cloud-native storage environment, particularly in the context of handling complex data types such as tensors. By improving the efficiency of tensor storage, the performance of machine learning applications can be enhanced, contributing to the broader field of artificial intelligence.

## VII. FUTURE WORK

While the current study has made progress toward optimizing tensor storage in cloud object storage environments, several areas for further exploration and refinement remain:

- **Cloud-native Enhancement:** The current query engine operates locally due to cost considerations, limiting the network bandwidth to 1Gbps. The use of an EC2 cluster from Amazon Virtual Private Cloud (Amazon VPC) could potentially increase this bandwidth to 100 Gbps. Therefore, future work should investigate the write and read performance of the proposed methods in such high-bandwidth environments while keeping other factors constant.

- **More Datasets Benchmarking:** The study's scope was limited to two datasets. An expansion to larger and more complex datasets will provide a more comprehensive understanding of the storage and retrieval performance of the proposed methods in real-world scenarios.

- **Overhead Reduction:** The current need for Spark to interact with Delta Lake introduces overhead, and some operations still rely on Python, which is less efficient for tensor operations. Future work will aim to implement a tensor data type directly in Delta Lake and use Ray, a distributed computing framework [29], for interactions, thereby reducing overhead and improving efficiency.

- **Parquet Configuration Analysis:** Delta Lake uses Parquet [22], and the experiments have relied on the default Parquet configurations. Future studies should perform a fine-tuned and quantified trade-off analysis of different Parquet configurations for Delta Lake tables to optimize storage and retrieval performance further.

- **Integration with ML Frameworks and vector databases:** The proposed methods' integration with popular machine learning frameworks, such as PyTorch [28], remains under exploration. Moreover, the current indexing uses tensor ID, but future work could explore vector indexing, such as similarity search and approximate nearest neighbor (ANN), to enhance data retrieval efficiency.

- **Energy Efficiency Analysis:** Given the significant energy consumption of data centers, an analysis of the energy efficiency of the proposed methods will be a valuable addition to future work. This analysis will contribute to the broader goal of sustainable computing, aligning technological advancements with environmental considerations.